\documentclass{article}

\usepackage{comment}
\usepackage{PRIMEarxiv}

\usepackage[utf8]{inputenc} 
\usepackage[T1]{fontenc}    
\usepackage{hyperref}       
\usepackage{url}            
\usepackage{booktabs}       
\usepackage{amsfonts}       
\usepackage{nicefrac}       
\usepackage{microtype}      
\usepackage{lipsum}
\usepackage{fancyhdr}       
\usepackage{graphicx}       
\graphicspath{{media/}}     
\usepackage[dvipsnames]{xcolor}
\usepackage[font=scriptsize,skip=1pt]{caption}
\usepackage{enumerate}
\usepackage{enumitem}
\usepackage{subcaption}
\usepackage{caption}
\usepackage[normalem]{ulem}
\usepackage[numbers, sort&compress, comma, square]{natbib}
\usepackage{amsmath}

\DeclareMathOperator*{\argmin}{arg\,min}

\usepackage{multirow}
\title{A Systematic Evaluation of Euclidean Alignment with Deep Learning for EEG Decoding}

\author{Bruna Junqueira$^{1, 2}$ \quad Bruno Aristimunha$^{2, 3}$ \quad \textbf{Sylvain Chevallier}$^{2}$ \quad Raphael Y. de Camargo$^{3}$ \\
\\
$^1$University of São Paulo, Sao Paulo, Brazil\\
$^2$Université Paris-Saclay, Inria TAU team, LISN-CNRS, France \\ $^3$Federal University of ABC, Santo Andre, Brazil \\ 
Corresponding author: brunaafl@usp.br
}

\begin{document}
\maketitle

\begin{abstract}
Electroencephalography (EEG) signals are frequently used for various Brain-Computer Interface (BCI) tasks. While Deep Learning (DL) techniques have shown promising results, they are hindered by the substantial data requirements. By leveraging data from multiple subjects, transfer learning enables more effective training of DL models. A technique that is gaining popularity is Euclidean Alignment (EA) due to its ease of use, low computational complexity, and compatibility with Deep Learning models. However, few studies evaluate its impact on the training performance of shared and individual DL models.
In this work, we systematically evaluate the effect of EA combined with DL for decoding BCI signals. We used EA to train shared models with data from multiple subjects and evaluated its transferability to new subjects. Our experimental results show that it improves decoding in the target subject by 4.33\% and decreases convergence time by more than 70\%. We also trained individual models for each subject to use as a majority-voting ensemble classifier. In this scenario, using EA improved the 3-model ensemble accuracy by 3.71\%. However, when compared to the shared model with EA, the ensemble accuracy was 3.62\% lower.
\end{abstract}


\section{Introduction}

Electroencephalography (EEG) signals are the most common type of neurophysiological data used for decoding tasks in Brain-Computer interfaces (BCI). It is non-invasive, portable, and has a low acquisition cost, unlike other technologies, such as magnetoencephalography (MEG) and functional magnetic resonance imaging (fMRI), which are expensive and require large equipment, and electrocorticography (ECoG), which require surgical procedures. Despite those advantages, EEG data also presents some drawbacks. It has a low signal-to-noise ratio (SNR), is vulnerable to the position of the electrodes in the head, and has a poor spatial resolution. 

Deep Learning (DL) algorithms have shown promising results in the BCI field, in some cases outperforming traditional machine learning methods for EEG classification~\citep{Schirrmeister2017, DOSE2018532, Lawhern2018, Roy2019}. Despite the promising results of DL, it's important to acknowledge that DL models are data-hungry~\citep{lecun2015deep}. Existing efforts predominantly focus on either training a single model for a group of subjects (\emph{Cross-Subject}, or \emph{Inter-Model}) or training an individual model for each subject (\emph{Within-Subject}, or \emph{Intra-Model})~\citep{CNN_REVIEW, Schirrmeister2017, Roy2019}. 

A way to improve the performance of DL models would be to use data from multiple subjects. Along with the expected increase in accuracy, it could reduce the calibration time to avoid unnecessary cognitive fatigue from the user~\cite{khazem2021minimizing}. But brain signals depend on the morphology and psychological state of the individual at the instant of the recording~\cite{Lotte2018update}. Therefore, variabilities of a person's signals between different moments in the same or between sessions make the decoding process more challenging. When considering different subjects, using models trained in one subject for decoding another can even worsen the predictions, a phenomenon called \emph{negative transfer}~\cite{Pan2010}. These signal variabilities are one of the main factors causing the generalization problem in the BCI field, which is exacerbated by the relatively small size of EEG-based datasets.

This is the problem targeted by the field of Transfer Learning (TL)~\cite{transferEEG, aristimunha2023evaluating, guetschel2023transfer}, which aims to improve the transferability of models between domains, where each domain can be an individual in a given task. There are two main approaches for TL: task and domain adaptation \cite{Pan2010}. In task adaptation, one trains the model in a given domain and, when applying it to the target domains, fine-tune the model parameters to improve the performance on the target domain. In domain adaptation, we use the same model in both domains but change the input data so that both domains have similar representations.

When considered in the BCI application, a task adaptation would be to train the model on another subject and \emph{fine-tune} the model using labeled data from the target~\cite{Yosinski2014freezeft,kalunga2018transfer}. When using domain adaptation, one would apply pre-processing transformations to make the source and target EEG signals more similar~\cite{Wei2021, Zhang2021}. One well-known technique is the Riemannian Alignments (RA) ~\citep{rodrigues18:procrutes, Zanini18:RAMDRM, Lotte15:RA, bleuze2022tangent}, which aims to reduce the distance between the subjects' covariance matrices in the Riemannian manifold~\cite{Riemannian2017}. For instance, \citet{rodrigues18:procrutes} propose a method based on the Procrustes analysis, applying geometrical transformations (re-centering, re-scale, rotation) on the Riemannian manifold. On the other hand, \citet{Zanini18:RAMDRM} insert paradigm information trying to re-centering based on the resting state trials. However, RA transformations are done on covariance matrices, which are incompatible with standard state-of-the-art DL models that operate on raw EEG data. There have been efforts to apply DL networks that work on covariance matrices, such as SPDNets \citep{huang2016riemannian} to EEG decoding \citep{wilson2023deep, Carrara2024GeometricNN} and Attention models to detect cognitive impairements~\citep{Qin2024BNMTrans}, but the number of studies is still limited.

Another domain adaptation method is the Euclidean Alignment (EA)~\citep{He2020:euclidean}, which works directly on the EEG signal. This form of data alignment reduces the distance between data distributions by centering their Euclidean mean covariance matrix to the identity. As a result of its simplicity and computational efficiency, there has been an increased interest in using EA for EEG decoding, usually as a pre-processing step combined with state-of-the-art models \citep{wei20222021, Kostas2020ThinkerIE, mi13060927}, and has been showing some good results on both traditional machine learning~\citep{Demsy2021, Li2021} and deep learning pipelines \citep{Dongrui2022} and in the context of merging datasets ~\citep{Ouahidi2023Interp, Mellot2024PhysicsinformedAU}. Despite its growing popularity, there are limited studies on how EA can affect deep learning models' training and transfer performance for shared models trained with data from multiple subjects and individual models trained with data from single subjects. 



Deep learning models are the main beneficiaries of transfer learning approaches since they benefit from large amounts of data. However, most existing DL approaches that use data from multiple subjects tend to ignore or overlook the issue of inter-subject variability \citep{farahat2019cnn, BORRA2023107323, Borra2021, Borra_2022, KHADEMI2022105288, Yap2023, Xu2021}, probably with the expectation that the DL model could accommodate for these differences. For this reason, it is essential to evaluate if applying data alignment approaches as a preprocessing step can improve the learning and inference processes. Moreover, it is crucial to closely examine the alignment mechanism's intricacies, its impact on the training set, and its influence on transferability across subjects when employing the EA. Other researchers had already compared the use of the Euclidean Alignment with deep learning models~\citep{Dongrui2022, Kostas2020ThinkerIE, MIAO2023120209, ouahidi2023strong}, but in a very concise manner. \citet{Dongrui2022}, in their review, have compared the use of the EA in different Machine Learning and some DL models. In the DL case, however, they just considered the offline cross-subject models. 

In this work, we make a systematic evaluation of the effect of EA combined with deep learning models for decoding BCI signals. We used EA to train shared models with data from multiple subjects and evaluated its transferability to new subjects. Our experimental results show that it improves decoding in the target subject by 4.33\% and decreases convergence time by more than 70\%. We also applied EA to models trained from individual source subjects and evaluated the improvements in transferability to target subjects, both for a single model and for majority-voting classifiers. In this scenario, using EA improved the 3-model classifier accuracy by 3.71\%. However, when compared to the shared model with EA, the majority-voting classifier accuracy was 3.62\% lower.

\section{Methodology}
\label{sec:method}

This section provides a sound and complete explanation of the Euclidean Alignment method and describes the datasets used and the different training schemes proposed. 

\subsection{EEG Decoding}

EEG decoding is the problem of classifying EEG signals, for instance, to retrieve mental states from the recorded signals~\citep{encodingdecoding}. We can define dataset $E$ as the sample of $N$ pairs of trials and labels,  $E=\{(\textbf{x}_i, y_i)\}_{i=1}^N$. The trial recordings $\textbf{x}$ are matrices having $c$ rows and $t$ columns, where $c$ is the number of channels used for the sampling process and $t$ is the number of time steps. Therefore, the feature space is $\mathcal{X}\in\mathbb{R}^{c \times t}$. In our experiments, we deal solely with the left-right imagery paradigm, and the decoding labels space is $\mathcal{Y}=\{\mathrm{'left\_hand'}, \mathrm{'right\_hand'}\}$, encode as $\{0, 1\}$, respectively. We work with data from more than one subject, and we consider each subject ($k$) to have their own data $E_k=\{(\textbf{x}^{k}_{i},y^{k}_{i})\}_{i=1}^{n}$, with $\left\{\mathcal{X}_k, \mathcal{Y}_k\right\} \in \mathbb{R}$. 

Because of brain signal variability, we consider that the marginal distributions $P_k(\cdot)$ of EEG data $E$ are different from subject to subject. The classification hypothesis assumes that there exists some unknown decision function $f_{\theta}(\cdot)$ with $\theta$ parameters that result in $f_{\theta}(\textbf{x})=y$. We are interested in optimizing the $\theta$ parameters to allow us to decode non-labeled brain signals. This is done by feeding a neural network with some labeled data (training $t$ data formed by $N_{t} < N$ examples) and minimizing the average loss $\ell$:

\begin{equation}
    \min_{\theta} \frac{1}{N_{t}}\sum_{i=1}^{N_{t}}{\ell(f_{\theta}(\textbf{x}_i),y_i)}
\end{equation}

In all our experiments, we used the negative log-likelihood loss ($\ell$) and AdamW optimizer for the minimization step.

\subsection{Transfer Learning}

As defined in \citep{zhuang2020comprehensive}, a \emph{domain} $\mathcal{D}$ is the set formed by the feature space $\mathcal{X}$ and a marginal distribution $P(\textbf{X})$, 
 with $\textbf{X}=(\textbf{x}_{1}, \dots, \textbf{x}_{N})$. The \emph{task} $\mathcal{T}$ is composed of the label set $\mathcal{Y}$ and the decision function $f$, which we want to discover. Mathematically, $\mathcal{D}=\{\mathcal{X}, P($\textbf{X}$)\}$ and $\mathcal{T}=\{\mathcal{Y}, f($\textbf{X}$)\}$. We consider each person a \emph{domain}, and we call the source domain the domain with known knowledge, and the target is the unknown one. Transfer learning is the framework by which, given a source domain ($\mathcal{D}_s$), source task ($\mathcal{T}_s$), and target domain ($\mathcal{D}_t$),  target \emph{task} ($\mathcal{T}_t$), with $\mathcal{D}_s \neq \mathcal{D}_t$ \emph{or} $\mathcal{T}_s \neq \mathcal{T}_t$, the knowledge from the source ($s$) is used to learn the target's ($t$) (knowledge).

Since we deal with data from different subjects, we can define a domain as $\mathcal{D}_k=\{\mathcal{X}_k, P_k(\textbf{X}^k)\}$ and a task $\mathcal{T}_k=\{\mathcal{Y}_k, f_{\theta}^k($\textbf{X}$^k)\}$ for each individual $k$. Given of the nature of the data, it is reasonable to consider that feature space $\mathcal{X}_k$, label space $\mathcal{Y}_k$, and decision function $f_{\theta}^k(\cdot)$ share a same attribute space for each subject $k$, but, the marginal distributions $P_k(\cdot)$ of EEG data are different from subject to subject. We use a \emph{leave-one-subject-out} approach, with multiple sources ($k-1$ subjects) with different marginal distributions $P(\cdot)$, where we approximate the marginal distributions using a data alignment technique called \emph{Euclidean Alignment}.



We also evaluated the transferability between every pair of subjects, using a model trained in a source subject $f_{\theta}^s(\cdot)$ to decode the signal on all other subjects $t$. We define the mean accuracy of the $f_{\theta}^s(\cdot)$ to classify the EEG signals of other subjects as the \emph{transferability} of the model. Similarly, we denote as the \emph{receivability} of the model as the mean accuracy when applying the individual models $f_{\theta}^s(\cdot)$ from all source subjects $s$ to the EEG data from the target subject.


\subsection{Euclidean Alignment}\label{sec:eu_al}




The inherent variability in brain signals poses a significant challenge in developing a model that can effectively generalize across diverse individuals.
To address this challenge, we analyze the effectiveness of Euclidean Alignment (EA)~\cite{He2020:euclidean}, which matches each subject's mean covariance matrix of EEG trials with the identity matrix, reducing their dissimilarities~\cite{Gretton2008:MMD}. 

Suppose that, for each of the $N$ subjects, we have a set of $n$ trials, $\textbf{x}_1, \textbf{x}_2, ..., \textbf{x}_n$. For each subject $j$, we first calculate the arithmetic mean $\bar{R^j}$ of the covariance matrices of all trials. 

\begin{equation}
    \bar{R}^j=\frac{1}{n} \sum^{n}_{i=1}X^j_i {X^j_i}^{\top},
\end{equation}

where $X^{j}_{i}$ represents the covariance matrix for the \textit{i}th EEG trial, of the \textit{j}th subject. Afterward, we define the transformation matrix as the square root of the inverse of $\bar{R^j}$ and apply it to each trial of the subject: 

\begin{equation}
\tilde{X^j_i}=({\bar{R}^j})^{-1/2} X^j_i.
\end{equation}

\begin{figure}[!ht]
    \centering
    \includegraphics[width=0.9\linewidth]{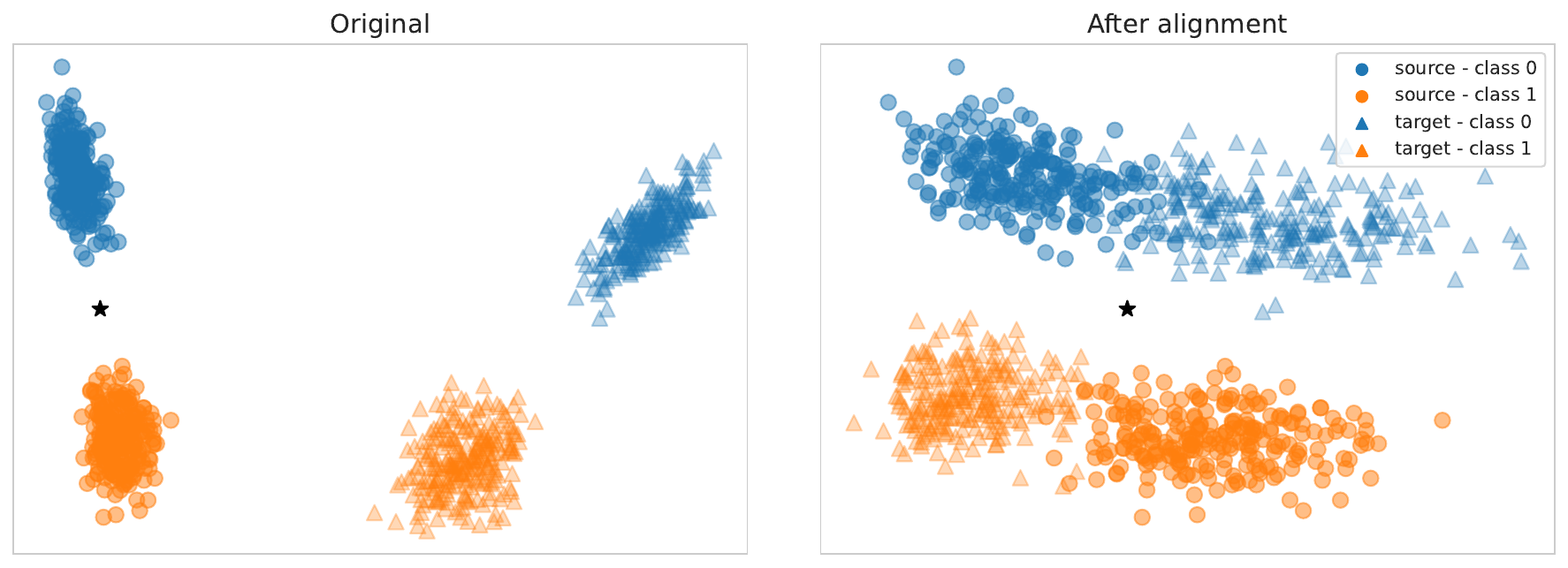} 
    \caption{Euclidean Alignment steps illustrated. The star represents the origin of the space. }
    \label{fig:example_eucl_align}
\end{figure}

As a result, the arithmetic mean of the covariance matrices of each subject becomes equal to the identity matrix, making the distributions more similar, as illustrated in \autoref{fig:example_eucl_align}.

\subsection{Riemannian Alignment}\label{sec:re_al}

Riemannian Alignment (RA)~\citep{congedo2017riemann} is another set of transformations to align EEG signals, but applied to the covariance matrices of EEG signals. Some of these transformations, such as the complete Riemannian Procrustes (RPA)~\citet{rodrigues18:procrutes}, are difficult to adapt for use on raw EEG data. Consequently, we decided to use the first step of the RPA, which is the recentering to the geometric mean. This method takes advantage of the congruence invariance property~\citep{congedo2017riemann} of the Affine-Invariant metric to recenter the mean of a distribution to the Identity matrix. Let $M_R^{j}$ be the geometric mean of the covariance matrices $\Sigma_1, \Sigma_2, ..., \Sigma_n$ of subject j

\begin{equation}
    M_R^{j} = \argmin_{C \in \mathcal{P}}\sum_{i=1}^{n}\delta^{2}(C, \Sigma_i^{j}),
\end{equation}

with $\mathcal{P}$ the space of $c \times c$ SPD matrices and $\delta$ the Affine-Invariant distance. Since $(M_R^{j})^{-1/2}$ is also SPD, by the invariance property,

\begin{equation}
    \delta(\Sigma_i, M_R^{j}) = \delta((M_R^{j})^{-1/2}\Sigma_i(M_R^{j})^{-1/2},(M_R^{j})^{-1/2} M_R^{j} (M_R^{j})^{-1/2}) = \delta((M_R^{j})^{-1/2}\Sigma_i(M_R^{j})^{-1/2},I).
    \end{equation}

Therefore, the recentered matrix is obtained by $(M_R^{j})^{-1/2}\Sigma_i (M_R^{j})^{-1/2}$. To maintain the data in the time domain, we applied the geometric mean $\bar{R}_{R}$ directly to the EEG signal: $\bar{R}_{R}^{-1/2}X$. 

The Riemannian Procrustes has other steps, but those need to be applied directly to the covariance matrices. \citet{He2020:euclidean} used the recentering to the resting state \citep{Zanini18:RAMDRM} instead of the geometric mean of trials to compare to EA. However, we consider that the geometric mean is more similar to the steps performed by the Euclidean Alignment.

\subsection{Fine-Tuning} 

We used linear probing, a sub-type of fine-tuning. We define a neural network $f_{\theta}:\mathcal{X}\rightarrow \mathcal{Y}$ as a composition of different functions $f_{\theta}=f_{\theta_1} \circ \dots \circ f_{\theta_l}$, where each function $i$ have $\theta_i$ corresponding to parameters from a subpart of the model. In the fine-tuning, we replace the last layer by $f_{\phi}$, resulting in $f'_{\theta'}=f_{\theta_1} \circ \dots \circ f_{\theta_{l-1}} \circ f_{\phi}$, and update only the weights of this last layer, freezing the others.

\subsection{Datasets}

We used two datasets available at the \textsc{MOABB}~\cite{Aristimunha_Mother_of_all_2023} repository. The first is the dataset IIa from BCI Competition 4 \cite{BNCI2014001} (\emph{BNCI2014}).
It consists of data from 9 healthy individuals, where each individual can perform four different motor imagery tasks: left hand (class 1), right hand (class 2), both feet (class 3), and tongue (class 4). It also has two sessions, recorded on different days, named train session (\emph{session$\_$T}) and evaluation session (\emph{session$\_$E}). Each session contains six runs, separated by a \emph{short break}, and each run contains 12 trials per class. We used both sessions, but only the first two classes, following other BCI studies. The EEG signals contain 22 channels sampled with a frequency of 250 Hz.

The second dataset is the High-Gamma~\citep{Schirrmeister2017}, denoted here as \emph{Schirrmeister2017}, containing data from 14 healthy individuals performing motor execution tasks: left hand, right hand, both feet, and rest. Just as it was done with BNCI2014, we used solely left- and right-hand tasks to maintain results comparable. Data is divided into two sessions:  training, with approximately 880 trials, and testing, with approximately 160 trials. The EEG trials contain 128 channels, sampled at 512 Hz. 




We employed standard pre-processing steps designed for the \emph{Motor Imagery} (MI) based in \citep{nam2018brain,Aristimunha_Mother_of_all_2023}. We applied band-pass filtering with the overlap-add method between $[8-32]$\ Hz and resampled the signals to $250$Hz. Additionally, we selected the 22 channels based on \citet{BNCI2014001} to standardize with the BNCI2014 dataset. 

\subsection{Models and Data Splitting}
\label{model_split}

\begin{figure}[!ht]
\centering
\begin{subfigure}[t]{0.32\textwidth}
    \includegraphics[width=\textwidth]{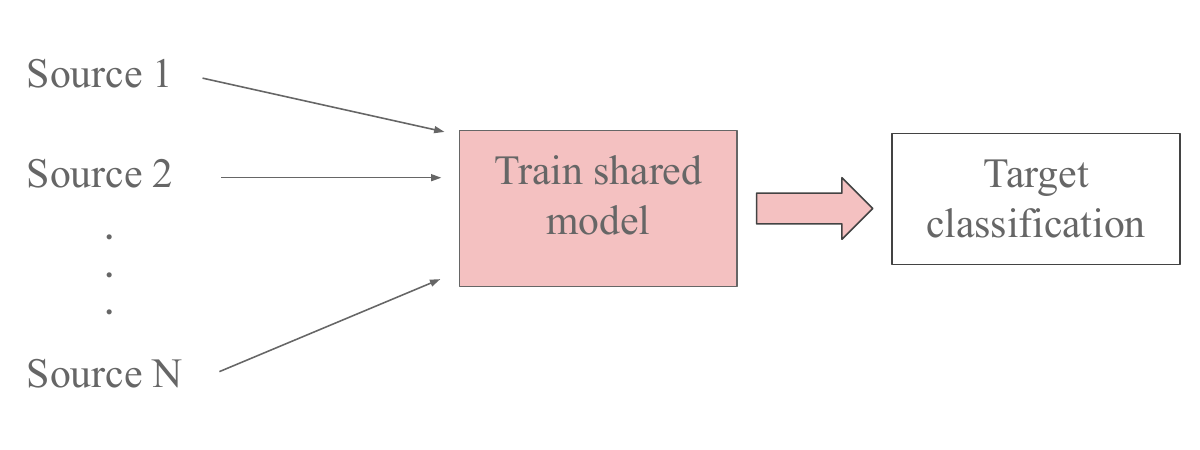}
    \caption{}
    \label{fig:slides_exp1}
\end{subfigure}
\quad
\begin{subfigure}[t]{0.3\textwidth}
    \includegraphics[width=\textwidth]{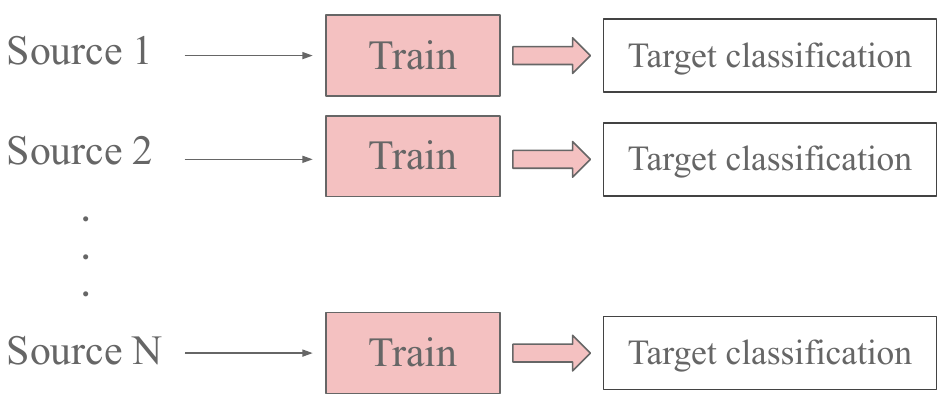}
    \caption{}
    \label{fig:slides_exp4a}
\end{subfigure}
\quad
\begin{subfigure}[t]{0.32\textwidth}
    \includegraphics[width=\textwidth]{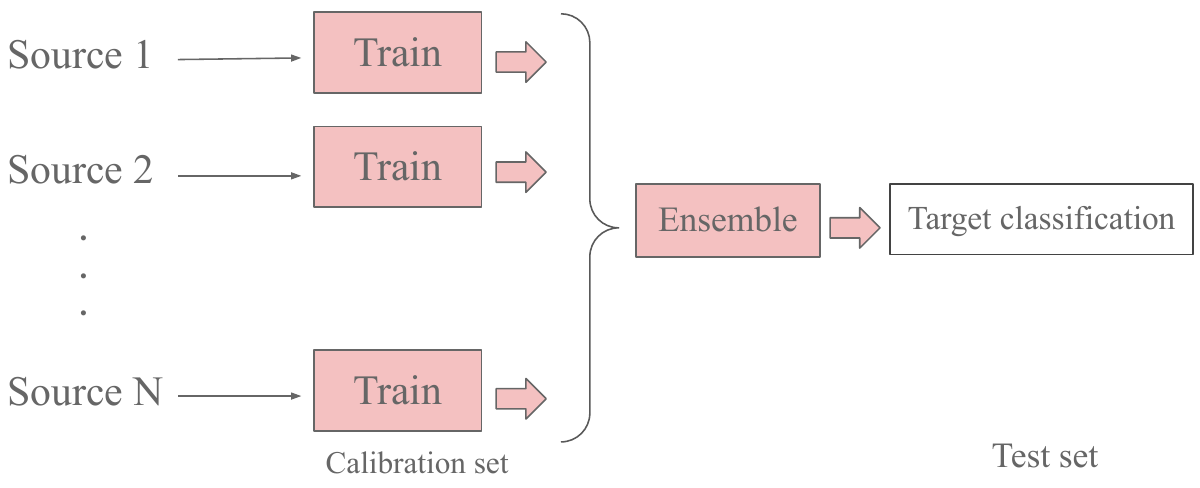}
    \caption{}
    \label{fig:slides_exp4b}
\end{subfigure}
\caption{Shared, individual, and majority-voting classifier models. (a) We use data from all source subjects to train a single shared model, with and without EA. (b) We use data from each source subject to train a separate individual model, with and without EA. (c) We use calibration data from the target subject to select the best source models and combine them in a majority-voting classifier.
} \label{fig:models}
\end{figure}

We analyzed the use of EA in two different scenarios: 

\begin{enumerate}[label=\roman*.]
    \item Using Cross-Subject models (\emph{\textbf{shared}}) trained with data from all source subjects (Figure~\ref{fig:models}a). We relied on a Leave-One-Subject-Out (LOSO) validation strategy for data split: at each time, we picked one subject as the target (test set) while using the remaining subjects as the training data; 
    \item Using \emph{\textbf{individual}} models, one per source subject, trained with data only from that subject. We evaluated each model separately (Figure~\ref{fig:models}b) or combined in majority-voting classifiers (Figure~\ref{fig:models}c), using the other subjects as test subjects.
\end{enumerate}

We then use these models to perform two-class motor imagery classification in the target subjects. These methodologies were carefully chosen to ensure robust evaluations and unbiased results by preventing data leakage between test, validation, and training sets \cite{white2023k}.

For the analysis, we evaluated each scenario using \emph{offline} and \emph{pseudo-online} evaluation approaches \cite{carrara2023online, martin_online:2024}. For the offline one, we used all data from the target subject to determine the reference matrix for the Euclidean alignment, while in the pseudo-online, we used only the first group of $24$ trials from the target to compute the reference matrix. For the experiments with fine-tuning, we also used the first $24$ trials in the fine-tuning process. We selected 24 trials to match the length of a single run in the \textit{BNCI2014} dataset and by following the study conducted by \cite{martin_online:2024}, which recommends using between 16 and 32 trials for alignment.

We used EEGNet~\cite{Lawhern2018}, ShallowNet, and DeepNet~\cite{Schirrmeister2017} for the evaluations since they are the standard benchmarks in BCI tasks while showing good accuracy in BCI cross-subject classification tasks~\cite{Dongrui2022}. We used the model's implementations from \textsc{Braindecode}~\cite{Schirrmeister2017}. 

\subsection{Data alignment}

For the EEG alignment step, we applied different methods for the \emph{offline} and \emph{pseudo-online} experiments. For the offline approach, we had access to all trials for each subject, so we aligned each subject in groups of $24$ trials to reduce noise in the reference matrices for transformations \cite{Zanini18:RAMDRM}. Regarding the simulated online context, we aligned the target data using the first group of $24$ trials from the target subject to compute the reference matrix and then used this matrix to align the test data. In the BNCI2014001 dataset \cite{BNCI2014001}, there are $12$ trials per class per run, and as we use two events (left and right hand), we simply align each run.

However, the High-gamma dataset \cite{Schirrmeister2017} presented an additional challenge as it does not have a consistent number of trials per subject, and it may not always be a multiple of $24$. To address this, we performed an extra pre-alignment step where we randomly removed some trials while maintaining the class proportion, ensuring that the remaining trials were divisible by $24$. 

\subsection{Model Training}



We trained the \textbf{\emph{shared models}} using the leave-one-subject-out paradigm. We used all data from the source subjects since the models were evaluated in different target subjects. Inside the source subjects set, we initially randomly separated 20\% for validation and 80\% for training. For the shared models, we used a batch of $64$ trials with $1000$ epochs, an early stopping of $250$ iterations, and a dropout rate of $25\%$. For each architecture, we executed a Grid search using leave-one-group-out cross-validation to optimize the learning rate and weight decay, with the search space shown in \autoref{tabular:search}. We included a lower learning rate for shared models because they have more diversified data, which could lead to more instability during training. We ended up with $w_d = 1e^{-5}$ weight decay for all models and a learning rate of $\eta = 1e^{-3}$ for both EEGNet and DeepNet and $\eta = 8.25e^{-4}$ for Shallow ConvNet. When using fine-tuning, we retrained them for an additional $600$ epochs, with a patience of $150$. During fine-tuning, we used the first $24$ trials from each training subject and the first $24$ trials from the target.

For the \textbf{\emph{individual models}}, we also used all data for training, with 20\% for validation, since the models were evaluated on target subjects distinct from the source ones. We trained for $1000$ epochs, with a patient of 250 and a dropout rate of $35\%$. We executed a hyper-parameter tuning to select the best learning rate and ended up with $\eta = 8.25e^{-4}$ for both EEGNet and ShallowNet and $\eta = 7.25e^{-4}$ for DeepNet, as shown in \autoref{tabular:search}. For the individual models, we didn't execute fine-tuning, as it would be impractical in a real scenario to fine-tune each individual model to each target.

\begin{table}[!ht]
\centering
\caption{Search space of hyperparameters for the grid search procedure.}
\begin{tabular}{c|cc} 
\toprule 
  \multirow{1}{*}{\small \bf Hyperparameters } & \multicolumn{2}{c}{\bf \small Models}\\
    \cmidrule{2-3}
    \small ($1e^{-4}$) & \small Shared & \small Individual \\
  \midrule
\small \small Learning rate & \small [12.5, 10.0, 8.25, 6.25] & \small [12.5, 10.0, 9.25, 8.25, 7.25] \\
\small Weight decay & \small [0, 0.10, 1] & \small [0, 0.10, 1] \\
\bottomrule
\end{tabular}
\label{tabular:search} 
\vspace{-2mm}
\end{table}




\section{Results}
\label{sec:results}

We evaluated the effectiveness of Euclidean Alignment for the offline and pseudo-online scenarios, using the \emph{shared} and \emph{individual} models. For the shared models, we trained a separate model for each target subject using the leave-one-subject-out paradigm. For the individual models, we trained each network only once, as the majority-voting classifiers use different combinations of the individual models. We performed multiple experiments to answer the following research questions:

\begin{enumerate}
\item \textbf{EA impact:} Does EA improve a shared model transfer performance in the offline and pseudo-online contexts?
\item \textbf{Fine tuning:} Does fine-tuning improve the performance of shared models?
\item \textbf{Transferability:} How transferable are models learned on source subjects to a target subject?
\item \textbf{EA transferability:} Does EA improve the transferability of individual models across subjects?
\item \textbf{Invidividual vs shared models:} How do majority-voting classifiers compare to shared models when using EA?
\end{enumerate}

For individual models, we organized three different scenarios to evaluate the transfer learning performance and to choose the best classifier for the target:

\begin{enumerate}
\item \textbf{best-model(-EA)}: we use the first run of the target subject as a calibration set to select the individual model that provided the best accuracy and, in the EA case, to align the EEG data from the target subject;
\item \textbf{k-Models(-EA)}: we use the first run of the target subject as a calibration set to select the $k$ best models to be used as a majority voting classifier and, in the EA case, to align the EEG data from the target subject.
\end{enumerate}


We tested the significance of the results using a one-tailed permutation-based t-test using \textsc{MOABB's}~\cite{Aristimunha_Mother_of_all_2023} functions \textit{compute$\_$dataset$\_$statistics} and \textit{find$\_$significant$\_$differences} to compute the p-values for each pair of pipelines. The null hypothesis is that the performance of the pipelines is equal.




\subsection{Transfer-learning using Euclidean aligned shared models}

Euclidean Alignment improved the average decoding performance on all evaluated datasets and architectures, as shown in line No-EA, offline-EA, and online-EA in \autoref{table:shared_model}. The offline-EA had an accuracy of 1.26\% better than the online-EA, which occurs because all the target data is used to estimate the covariance matrices, resulting in a better alignment. However, the online-EA had an overall mean accuracy of 4.33\% better than the non-aligned case, showing the benefit of using EA in a more realistic scenario. Finally, using EA and RA resulted in very similar results, with slightly better accuracies for one of the approaches depending on the model or dataset. We use the Euclidean Alignment in the next experiments since it is the most used alignment method when combined with DL models.


\begin{table}[!ht]
\centering
\caption{Mean accuracy of different training schemes for the shared model using different datasets and pipelines.}
\resizebox{\textwidth}{!}{\begin{tabular}{l|ccc|cc|l} 
\toprule 
  \multirow{2}{*}{\bf Pipelines} & \multicolumn{3}{c|}{\bf Shared Models} & \multicolumn{2}{c|}{\bf Datasets} & \multirow{2}{*}{\bf Overall} \\
  \cmidrule{2-6}
  & EEGNet \cite{Lawhern2018} & DeepNet \cite{Schirrmeister2017} & ShallowNet \cite{Schirrmeister2017} & BNCI2014 \cite{BNCI2014001} & Schirrmeister2017 \cite{Schirrmeister2017} \\
  \midrule
No-EA & 72.18 $\pm$ 13.14 & 69.27 $\pm$ 12.44 & 69.99 $\pm$ 11.87 & 68.93 $\pm$ 12.61 & 72.46 $\pm$ 12.04 & 70.48 $\pm$ 12.43\\
Offline-EA & 76.62 $\pm$ 12.38 & 76.86 $\pm$ 11.82 & 74.72 $\pm$ 10.97 & 73.98 $\pm$ 11.21 & 78.75 $\pm$ 11.80 & 76.07 $\pm$ 11.66\\
Offline-RA & 76.42 $\pm$ 12.71 & 76.92 $\pm$ 11.93 & 74.23 $\pm$ 10.82 & 74.26 $\pm$ 11.11 & 77.91 $\pm$ 12.42 & 76.86 $\pm$ 11.78\\
\midrule
Online-EA & 74.55 $\pm$ 11.89 & \bf 75.92 $\pm$ 11.80 & \bf 73.95 $\pm$ 10.60 & 72.99 $\pm$ 11.53 & \bf 77.14 $\pm$ 10.81 & \bf 74.81 $\pm$ 11.35 \\
Online-RA & 75.25 $\pm$ 12.12 & 75.80 $\pm$ 12.52 & 72.12 $\pm$ 10.21 & 73.42 $\pm$ 11.59 & 75.63 $\pm$ 11.77 & 74.39 $\pm$ 11.66 \\
No-EA fine-tuning & 74.48 $\pm$ 12.35 & 70.74 $\pm$ 12.64 & 70.52 $\pm$ 11.13 & 70.70 $\pm$ 12.10 & 73.50 $\pm$ 11.99 & 71.91 $\pm$ 12.07 \\
EA fine-tuning & 74.85 $\pm$ 11.60 & 75.73 $\pm$ 11.40 & 73.79 $\pm$ 11.21 & 73.33 $\pm$ 11.35 & 76.67 $\pm$ 11.12 & 74.79 $\pm$ 11.31 \\
RA fine-tuning & \bf 75.65 $\pm$ 12.29 & 75.61 $\pm$ 12.08 & 72.39 $\pm$ 10.26 & \bf 73.75 $\pm$ 11.41 & 75.59 $\pm$ 11.80 & 74.55 $\pm$ 11.56 \\
\bottomrule
\multicolumn{7}{l}{1. {\bf Bold} for the best model between No-EA fine-tuning, Online-EA, Online-RA, EA fine-tuning, and RA fine-tuning. 
}
\end{tabular}}
\label{table:shared_model} 
\vspace{-2mm}
\end{table}

\begin{figure}[ht]
    \centering
    \includegraphics[width=\linewidth]{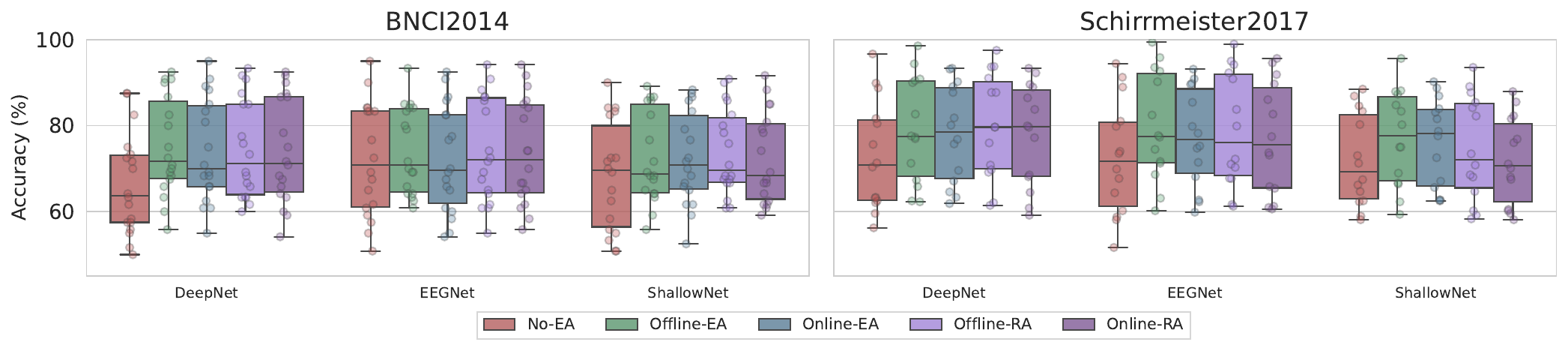}
    \caption{Prediction accuracy using ShallowNet, DeepNet, and EEGNet shared models without alignment, and with EA and RA, in the online and offline scenarios. Left: BNCI2014 dataset, right: Schirrmeister2017 dataset.}
    \label{fig:exp1}
\end{figure}

When considering the individual target subjects (Figure~\ref{fig:exp1}), there was no improvement in the median of subject accuracy in BNCI2014 for the EEGNet and ShallowNet, but the accuracy of the best and worst subjects improved in most cases, resulting in a better average accuracy. Also, comparing EA and RA resulted in similar patterns to those shown in \autoref{table:shared_model}, with better results for one of the approaches depending on the model or dataset. In the Schirrmeister dataset, the improvements were more evident, both in the median and average accuracies. This better result could be due to the increased number of source subjects to train the model, but we would need further investigation to evaluate this hypothesis. 

\begin{figure}[ht]
    \centering
    \includegraphics[width=\linewidth]{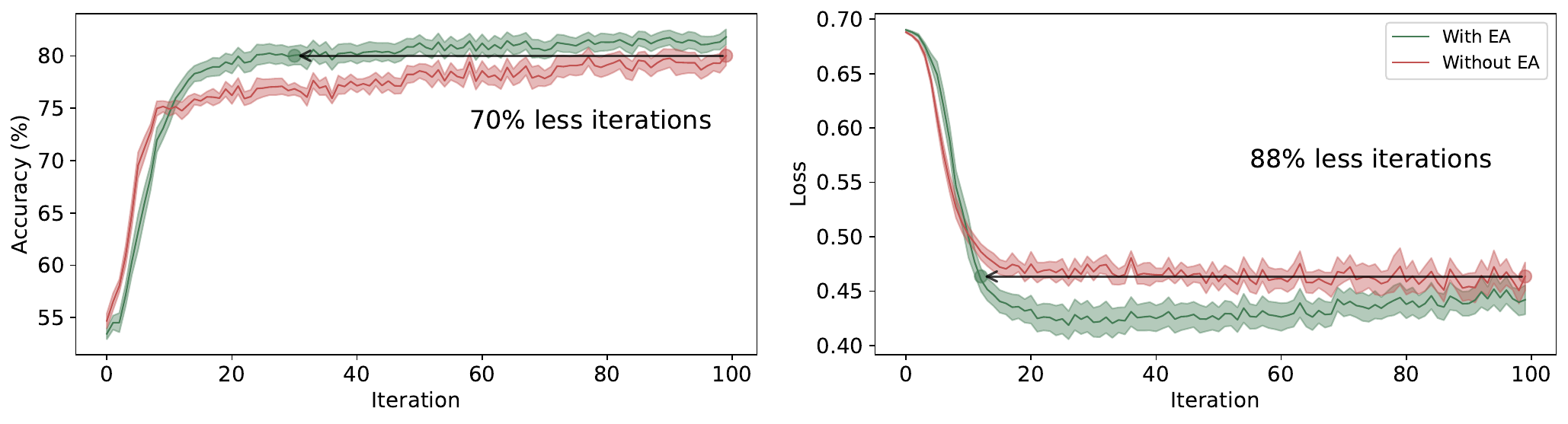}
    \caption{Validation accuracy and loss during training iterations using EEGNet and BCI2014 datasets, with shaded areas depicting the standard deviations over all trained models, one per target subject.}
    \label{fig:lc}
\end{figure}

There was also a significantly faster convergence in the training set in aligned models compared to their non-aligned counterparts, as illustrated in \autoref{fig:lc}. Specifically, within a predefined time window of 100 iterations, the aligned model achieved the same mean accuracy in $70\%$ fewer iterations as the non-aligned model and the same validation loss in just $88\%$ of the iterations.



\subsection{Fine-tuning the shared model}

Using the same amount of calibration data for the non-aligned pipeline, performing fine-tuning in the aligned models did not improve the mean accuracy, with some slight increases in accuracies in some datasets or models and decreases in others, as shown in Table~\ref{table:shared_model}. This may indicate that applying EA adjusts the domains in a way that, for simple fine-tuning to improve performance, it would need more target data. For the non-aligned case, fine-tuning leads to improvements in accuracy for all datasets and models, with an increase of 1.43\% in the general mean accuracy. When considering the median, there were no changes in all cases. However, for the non-aligned scenarios, the first and third quartiles improved with fine-tuning in most cases, resulting in improved average accuracies.


\begin{figure}[ht]
    \centering
    \includegraphics[width=0.6\linewidth]{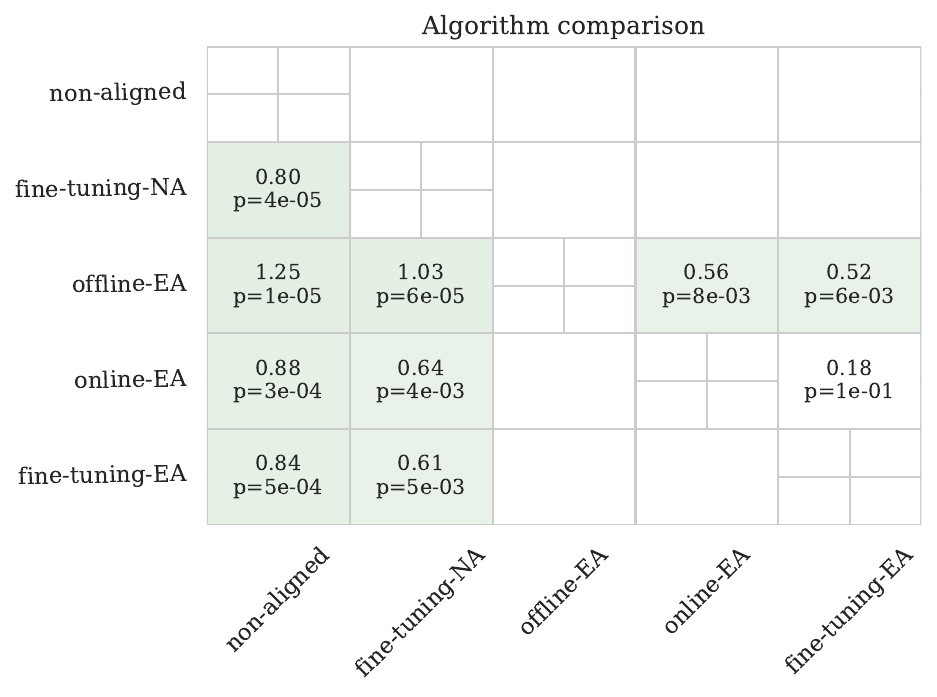}
    \caption{Significance matrix comparing all shared pipelines, with colored cells indicating a significant difference. The algorithms in the x-axis are the baselines, and the values shown correspond to the standardized mean difference between pipelines in the y-axis and the respective baselines and the p-values.}
    \label{fig:stat1}
\end{figure}


To test the significance of these shared model results, we executed a one-tailed permutation-based paired $t$-test (\autoref{fig:stat1}), as explained before. Using EA improved the classification in all tested scenarios, with the offline scenario significantly better than all other scenarios. Also, online EA was significantly better than all non-aligned scenarios, while fine-tuning improved the results significantly only for the non-aligned case.

\begin{figure}[ht]
    \centering
    \includegraphics[width=\linewidth]{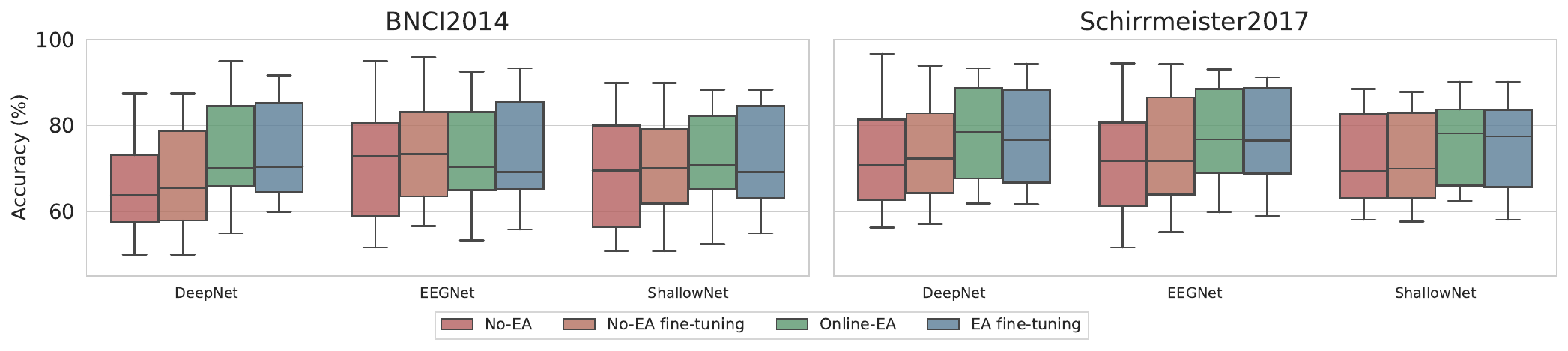}
    \caption{Accuracies of fine-tuned models using ShallowNet, DeepNet, and EEGNet shared models with and without EA.}
    \label{fig:exp2}
\end{figure}



\subsection{Analyzing transferability between subjects}

\begin{figure}[ht!]
  \centering  
  \begin{subfigure}{\textwidth}  
    \centering
    \includegraphics[width=\textwidth]{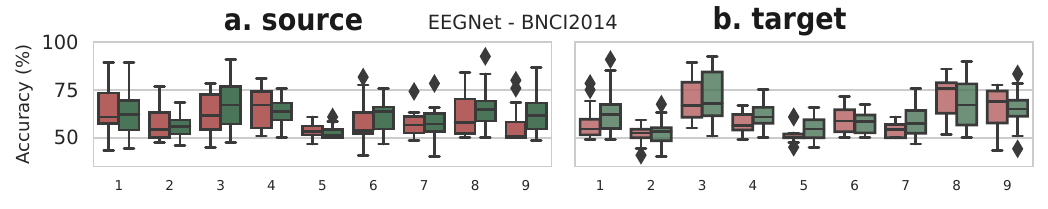}
    \includegraphics[width=\textwidth]{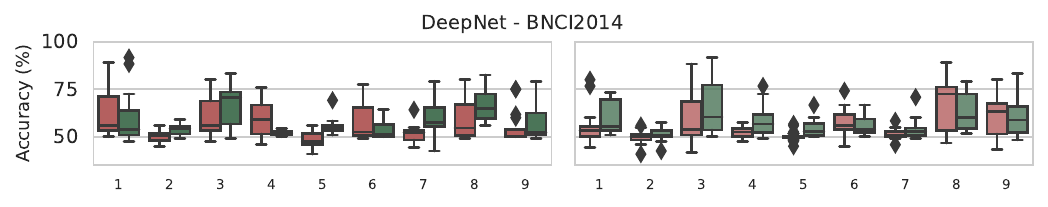}
    \includegraphics[width=\textwidth]{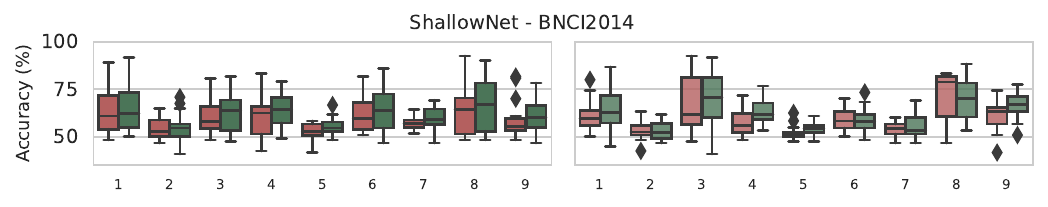}
    \includegraphics[width=\textwidth]{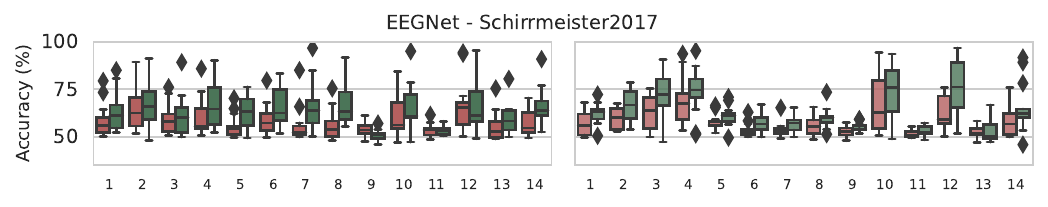}
    \includegraphics[width=\textwidth]{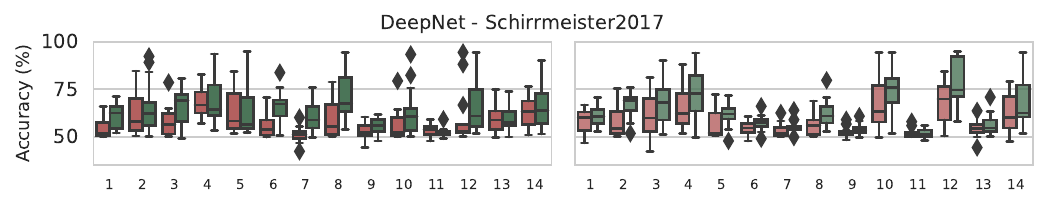}
    \includegraphics[width=\textwidth]{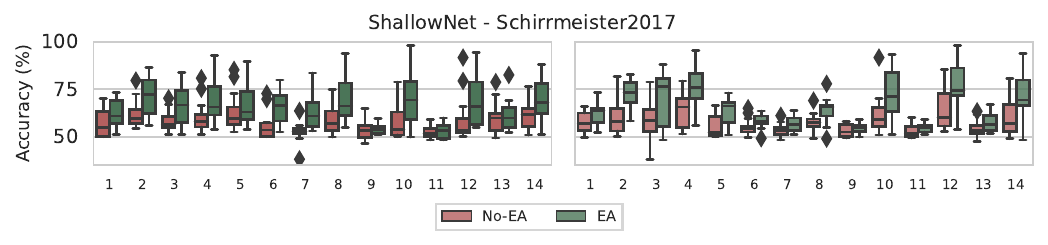}
  \end{subfigure}  
  
  \caption{Transferability of trained individual models for different datasets and neural network models. Graphs on the left show the transferability from each source model (a), while graphs on the right (b) show the receivability of target individuals to source models.}
  \label{fig:transferability}
\end{figure}

We evaluated how EA affects the transferability between subjects, evaluating the prediction accuracy per target and source subject. EA improved the transferability for most subjects, with an increase superior to 10\% in the median accuracy in several cases, as shown in \autoref{fig:transferability}.

\begin{figure}[ht]
  \centering  
  \begin{subfigure}{\textwidth}
  \centering       
    \includegraphics[width=\textwidth]{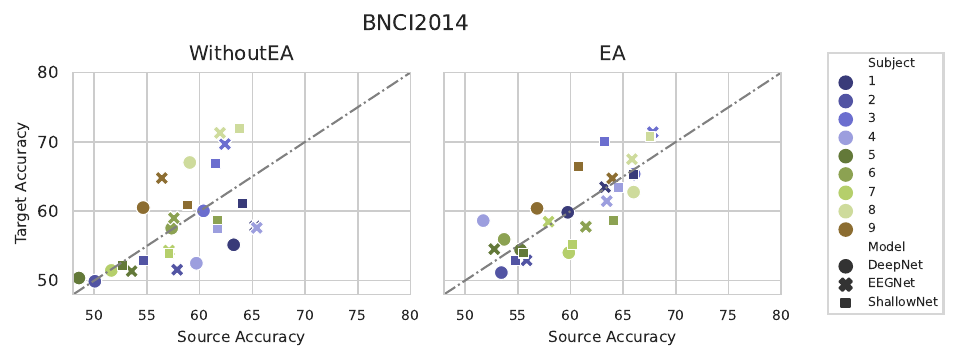}
    \caption{BNCI2014001 }    
    \label{fig:ensemble_eegneta}
    \end{subfigure}   
  \begin{subfigure}{\textwidth}
  \centering       
    \includegraphics[width=\textwidth]{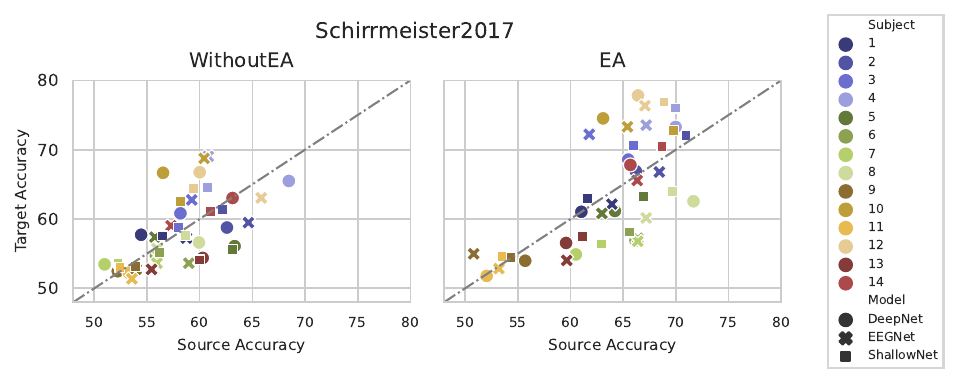}
        \caption{Schirrmeister }    
    \label{fig:ensemble_eegnetb}
    \end{subfigure}   
  \caption{Transferability plots with the relation between average source and average target accuracies for each subject on each model.}
  \label{fig:transfscatter}
\end{figure}

The analysis also indicates that, in general, individuals who are bad donors (sources), i.e., whose models, on average, do not have a good performance on different target subjects, also tend to be bad receivers (targets), and vice versa, as shown in \autoref{fig:transfscatter}. Applying Euclidean Alignment did not change the structure of good/bad donors and receivers but increased the correlation, as can be observed in \autoref{tabular:corr}. This behavior was expected since, in the aligned case, the data distribution becomes more similar between subjects. If the data from a source subject is useful for the target ones, we should expect the reverse to occur as well. Without alignment, data from different subjects are most likely spread in different areas in the domain space.

\begin{table}[ht!]
\centering
\caption{Mean Pearson correlation coefficient between the average accuracy of the subjects as sources and targets using individual models. The rows represent the models with and without alignment, and the columns indicate the different models and datasets.}
\resizebox{\textwidth}{!}{\begin{tabular}{l|ccc|cc|l} 
\toprule 
  \multirow{2}{*}{\bf Pipelines} & \multicolumn{3}{c|}{\bf Individual Models} & \multicolumn{2}{c|}{\bf Datasets} & \multirow{2}{*}{\bf Overall} \\
  \cmidrule{2-6}
  & EEGNet \cite{Lawhern2018} & DeepNet \cite{Schirrmeister2017} & ShallowNet \cite{Schirrmeister2017} & BNCI2014 \cite{BNCI2014001} & Schirrmeister2017 \cite{Schirrmeister2017} \\
  \midrule
No-EA & 0.41 & 0.50 & 0.49 & 0.52 & 0.41 & 0.47 \\
Offline-EA & 0.64 & 0.61 & 0.74 & 0.66 & 0.66 & 0.67 \\
\bottomrule
\end{tabular}}
\label{tabular:corr} 
\vspace{-2mm}
\end{table}

The transferability information is useful when selecting source individual models for the majority-voting classifiers since it may be more advantageous to choose only the best individuals for a specific target~\cite{ZHOU2002239}.


\subsection{Majority-voting classifiers for transfer learning}

Given the improvements in transferability, we evaluated the feasibility of using the best-transferred model or a majority voting over the best models to classify signals from the target subjects. We constructed the majority voting classifiers using the same small labeled set of target data from the online pipelines, using the exponential of the accuracy to define the weights from each model.


It is possible to notice from \autoref{table:ensemble} that, when training individual models, using EA also resulted in clear benefits, with a 3.71\% increase in the overall mean accuracy when comparing the 3-model classifiers with and without EA. The best mean accuracies occurred for all datasets and neural network models for the 3 and 5-model classifiers with EA.

\begin{figure}[!ht]
\centering
\begin{subfigure}[!t]{0.48\textwidth}
    \includegraphics[width=\textwidth]{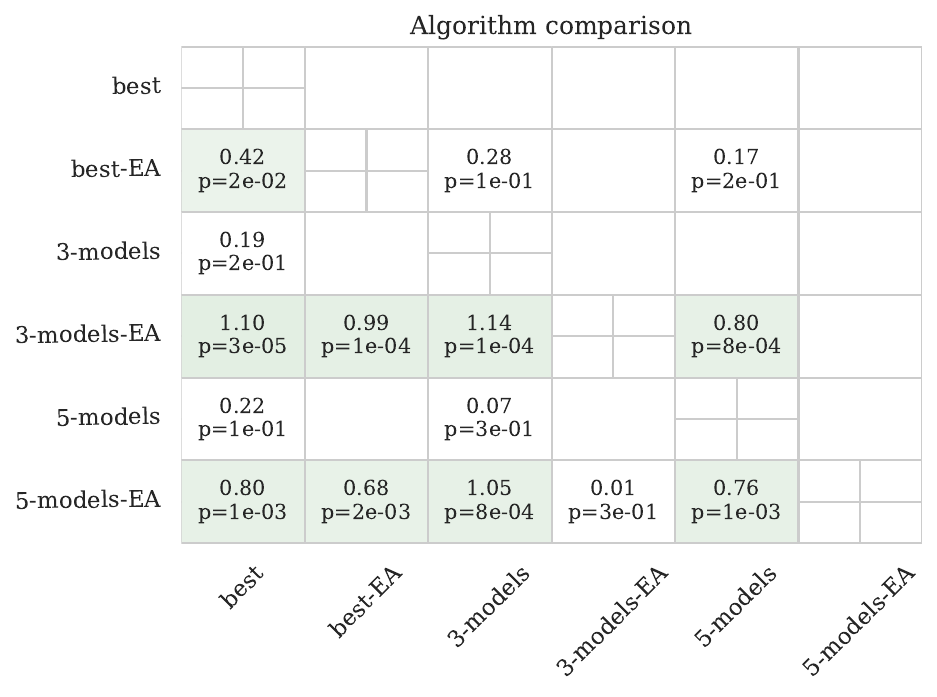}
    \caption{}
    \label{fig:stat_ensemble}
\end{subfigure}
\quad
\begin{subfigure}[!Ht]{0.48\textwidth}
    \includegraphics[width=\textwidth]{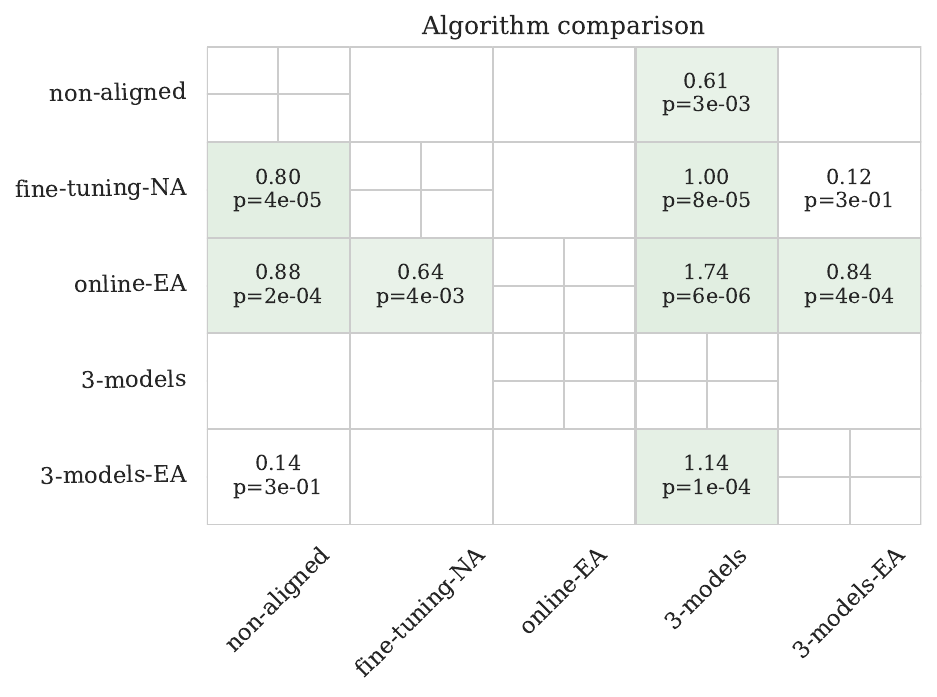}
    \caption{}
    \label{fig:stat_ensemble_shared}
\end{subfigure}
\caption{a) Significance matrix comparing all majority-voting pipelines. b) Significance matrix comparing the 3-model classifier and the shared model. Colored cells indicate a significant difference. The algorithms in the x-axis are the baselines, and the values shown correspond to the standardized mean difference between pipelines in the y-axis and the respective baselines and the p-values.}
 \label{fig:stat2}
\end{figure}

Moreover, in the non-aligned scenario, increasing the number of voter models did not increase the accuracy significantly compared to using only the best model~\autoref{fig:stat2}a. When using EA, classifiers with 3 and 5 models significantly improved accuracy compared to using the best model. This is probably because, without the alignment, very few source models have higher accuracies for each target, while with EA, more models transfer well. Although using the 3-model classifier produced a better mean accuracy than the 5-model classifier, the improvement was not statistically significant. We note that since we use a small number of trials to select the best models, our method will not always get the best models.

\begin{figure}[ht]
  \centering  
  \begin{subfigure}{\textwidth}
  \centering       
    \includegraphics[width=\textwidth]{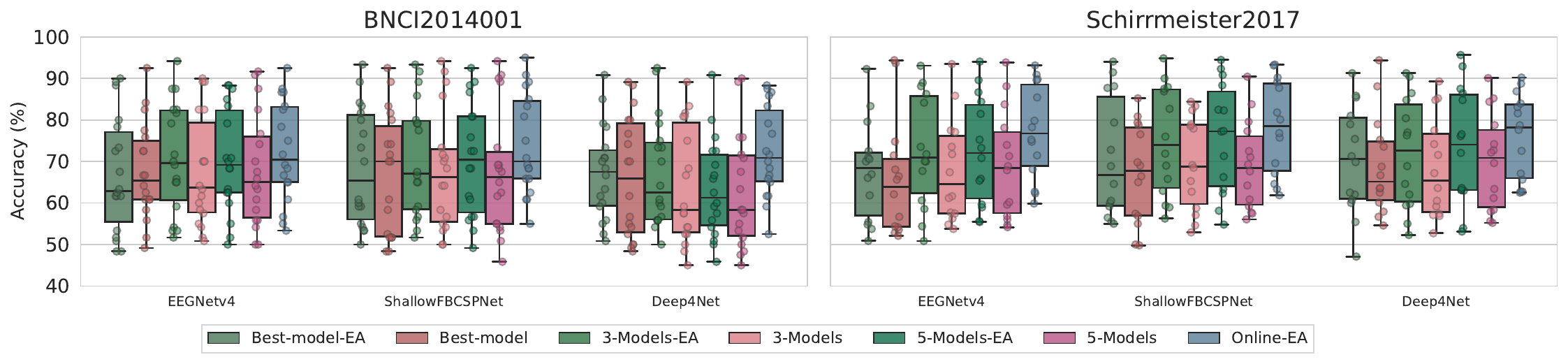}
    \end{subfigure}      

  \caption{Comparison between aligned and non-aligned majority-voting models test accuracies for the best model, 3 models, and 5 models, compared to the shared model with online EA.}
\label{fig:ensemble}
\end{figure}

When comparing the majority-voting classifier with the shared model (\autoref{fig:ensemble}), we can see that the shared model median accuracy is superior in all scenarios, with the biggest improvements when considering the lower quartile. This shows the highest benefit of using the shared model is for the subjects with the worst accuracy.

Finally, the shared model produced significantly better accuracies (\autoref{fig:stat2}b) when compared to majority-voting classifiers, with a difference of 3.62\% in the mean accuracy to the best majority-voting model. This indicates that training a neural network model using data from all subjects to adjust the weights is advantageous over majority-voting classifiers. On the other hand, it is easier to include more subjects when constructing new majority-voting models than in the shared one since the latter requires retraining the whole model.

\begin{table}[!ht]
\centering
\caption{Comparison between aligned and non-aligned majority-voting models test accuracies for the best model, 3 models, and 5 models.}
\resizebox{\textwidth}{!}{\begin{tabular}{l|ccc|cc|l} 
\toprule 
  \multirow{2}{*}{\bf Pipelines} & \multicolumn{3}{c|}{\bf Models} & \multicolumn{2}{c|}{\bf Datasets} & \multirow{2}{*}{\bf Overall} \\
  \cmidrule{2-6}
  & EEGNet \cite{Lawhern2018} & DeepNet \cite{Schirrmeister2017} & ShallowNet \cite{Schirrmeister2017} & BNCI2014 \cite{BNCI2014001} & Schirrmeister2017 \cite{Schirrmeister2017} \\
  \midrule
best-model & 66.72 $\pm$ 12.60 & 67.32 $\pm$ 13.35 & 67.19 $\pm$ 13.32 & 66.94 $\pm$ 13.47 & 67.25 $\pm$ 12.42 & 67.08 $\pm$ 12.96\\
3-Models & 67.79 $\pm$ 12.80 & 66.05 $\pm$ 13.25 & 68.60 $\pm$ 12.97 & 66.62 $\pm$ 13.85 & 68.59 $\pm$ 11.68 & 67.48 $\pm$ 12.92\\
5-Models & 68.33 $\pm$ 13.13 & 65.87 $\pm$ 13.18 & 68.18 $\pm$ 13.06 & 65.99 $\pm$ 13.99 & 69.35 $\pm$ 11.57 & 67.46 $\pm$ 13.03 \\
\midrule
best-model-EA & 67.22 $\pm$ 12.97 & 70.74 $\pm$ 11.95 & 69.70 $\pm$ 14.10 & 67.61 $\pm$ 12.97 & 69.74 $\pm$ 12.95 & 68.54 $\pm$ 12.94 \\
3-Models-EA & \bf 71.60 $\pm$ 13.33 & \bf 75.73 $\pm$ 13.19 & 72.65 $\pm$ 13.52 & \bf 69.58 $\pm$ 13.27 & 73.25 $\pm$ 13.18 &  \bf 71.19 $\pm$ 13.28 \\
5-Models-EA & 71.29 $\pm$ 12.62 & 75.73 $\pm$ 13.68 & \bf 72.85 $\pm$ 13.56 & 68.09 $\pm$ 12.91 & \bf 74.03 $\pm$ 13.23 & 70.69 $\pm$ 13.32 \\
\bottomrule
\multicolumn{7}{l}{1. {\bf Bold} for the best model between aligned versus non-aligned pipelines. 
}
\end{tabular}}
\label{table:ensemble} 

\end{table}

\section{Discussion}

Euclidean Alignment has emerged as a promising technique for domain adaptation in BCI systems, showing good results for transfer learning~\cite{He2020:euclidean}. Despite its promising results, most works have focused on using classical BCI decoding methods ~\citep{Demsy2021, Li2021}, and few works evaluated it in combination with Deep Learning methods ~\citep{Dongrui2022}. 

When using single shared models with training data from all individuals, using EA allowed us to achieve the same level of accuracy as the models without EA using 70\% fewer iterations and a final level of accuracy 4.33\% higher when using the same number of iterations. This faster convergence was not explored in earlier works and probably occurs because we provide data with more similar distributions among individuals. This is a point that can be further explored in future studies. 

Existing work using EA with Deep Learning focuses on the offline case, where all data from the target subject is used to determine the EA matrices. For instance, \citet{Dongrui2022} evaluates the use of offline EA in EEGNet and ShallowNet. However, in real applications, calibration trials should be reduced as much as possible to avoid early cognitive fatigue before starting the task. \citet{He2020:euclidean} evaluated the pseudo-online scenarios for the LDA algorithm and showed that 12\% of the target data is sufficient for a good estimation, but using more data improves the accuracy by a few percent. We demonstrated that this data frugality also applies to Deep Learning models. Using data from a single run with 24 trials of each class resulted in an accuracy of only 1.26\% below that of the offline case. This occurs because there is less data to determine the distribution of target classes in the latent space, and, unlike the offline case, the target data distribution differs from the data used to train the model. However, the online case is the only one that can be used in real scenarios.

Fine-tuning resulted in no improvements in accuracy for the shared model with EA and a 1.43\% increase in accuracy without EA. The small increase without EA possibly occurred because there was only a small amount of data to fine-tune the model. Using EA, since the models were aligned from the beginning, this small fine-tuning was insufficient to improve the accuracy. 

We also evaluated the scenario with multiple models trained with data from different subjects. One effect noted in other works applying RA is that subjects with good self-scores tend to be good ``receivers"~\cite{transcending}. Using EA, we did not evaluate the self-scores, but we noted that subjects are good ``donors" tend to be good ``receivers" and bad ``donors" tend to be bad ``receivers." We did not try to train the shared model using only the ``good'' donors, but this could be a way to improve the transfer accuracies.

When training individual models, using EA also resulted in clear benefits when selecting only the best source model and majority-voting classifiers of 3 and 5 models. Using weighted majority voting from 3 or 5 models worked better than just using the best model but had a 3.6\% reduction in the average accuracy compared to the online shared model. One advantage of the majority-voting models is that adding more source subjects to the collection of source models is easy. We only need to add the new source models to the selection process for the best transferable models for each target subject. One disadvantage is that the inference takes longer. However, it should be fast enough for real applications.

\citet{Kostas2020ThinkerIE} discusses the concepts of domain generalization (DG), where we adapt the source domains (subjects) to be useful for multiple target domains (subjects), such as when using EA, and domain adaptation (DA), where the domains are adapted to a single target subject. An example of DA would be selecting different source subjects to transfer to each target subject, as done in the majority-voting models. The authors argue that one can apply DG and DA concurrently and improve transfer learning, as we verified in our experiments with our majority-voting classifiers.

\subsection{Limitations}

It is important to address some of the limitations of our study. Firstly, we did not explore the impact of different ranges of hyperparameters, such as learning rate, in models with and without EA. For instance, a model with aligned input may accept larger learning rates since there would be less difference between batches. We also used the same hyperparameters for all individual models, but it is well-known that subject-specific hyperparameters can improve accuracy. Other interesting contributions would be to explore the faster convergence of aligned models better and analyze for improvements in the variance of predictions, using, for example, the Monte-Carlo Dropout technique.

One of the approaches employed in this study was to use a weighted majority voting mechanism over the best individual models. However, there are better ways to generalize knowledge across diverse subjects than combining models optimized for each individual. We selected this approach because the focus of the evaluation of individual models was to determine if Euclidean Alignment improves transferability between subjects.

\section{Conclusions}

We explored the impacts of the Euclidean Alignment, a domain adaptation technique commonly used on BCI pipelines for Transfer Learning, in the context of Deep Learning. We trained \emph{shared models} using data from multiple subjects and observed an absolute improvement of 5.55\% in the model accuracy for the offline case and 4.33\% in the pseudo-online case. Moreover, Euclidean Alignment improved the mean accuracy for all cross-subject models and datasets evaluated and led to a 70$\%$ acceleration in convergence in the shared models. Consequently, we believe that Euclidean Alignment should be a standard pre-processing step when training cross-subject models.





\section*{Acknowledgments}
BJ and RYC thank the São Paulo Research Foundation (FAPESP) for the financial support (grants 22/08920-0 and 23/06407-7). The work of BA was supported in part by the CAPES under Grant 001/003 and Data IA mobilité international,  Convergence Institute as part of the ``Programme d’Investissement d’Avenir'' (ANR-17-CONV-0003) operated by LISN.
\bibliographystyle{plainnat}  
\bibliography{references}

\end{document}